\documentclass[aps,pra,twocolumn,superscriptaddress]{revtex4-1}
\usepackage{ifthen,graphicx}
\usepackage{amsmath,amssymb,bm}
\usepackage[normalem]{ulem}

\newcommand{\figref}[1]{Fig.~\ref{fig:#1}}
\newcommand{\equref}[1]{Eq.~(\ref{eq:#1})}

\newcommand{\secref}[1]{Sec.~\ref{sec:#1}}
\begin{document}


\title{Perfect screening of the inter-polaronic interaction}


\author{Shimpei Endo}
\email[E-mail address: ]{endo@cat.phys.s.u-tokyo.ac.jp}
\affiliation{Department of Physics, University of Tokyo, 7-3-1, Hongo, Bunkyo-ku, Tokyo 113-0033, Japan}
\author{Masahito Ueda}
\affiliation{Department of Physics, University of Tokyo, 7-3-1, Hongo, Bunkyo-ku, Tokyo 113-0033, Japan}



\date{\today}

\begin{abstract}
We consider heavy particles immersed in a Fermi sea of light fermions, and study the interaction between the heavy particles induced by the surrounding light fermions. With the Born-Oppenheimer method, we analytically show that the induced interaction between $N$ heavy particles vanishes for any $N$ in the limit of high light-fermion density. The induced interaction vanishes even in the unitarity regime. This suggests that the formation of $N$-body bound states associated with the Efimov effect is suppressed in the presence of the dense Fermi sea. We ascribe the vanishing induced interaction to the screening effect in the neutral Fermi system.
\end{abstract}



\maketitle


\section{Introduction}

Ultracold atoms offer an ideal playground to study few-body and many-body physics in a controlled manner. By fine-tuning the $s$-wave scattering length $a_s$ between atoms using a Feshbach resonance~\cite{inouye1998observation,*chin2010feshbach}, three-body bound states called Efimov states~\cite{efimov1970energy,efimov1973energy,*braaten2006universality} have recently been observed through an enhanced atomic loss around the resonantly interacting regime $1/a_s=0$~\cite{kraemer2006evidence,*PhysicsF}. As for the many-body physics, the BEC-BCS crossover and its universal behavior around the resonant regime have been studied both experimentally~\cite{horikoshi2010measurement,*nascimbene2010exploring,*van2012feynman} and theoretically~\cite{PhysRevLett.92.090402,chen2005bcs,tan2008energetics,*tan2008large,*tan2008generalized}.

Polaron physics can also be studied with ultracold atoms~\cite{PhysRevLett.102.230402,*SalomonPolarColl,PhysRevLett.106.105301,*koschorreck2012attractive,kohstall2012metastability}. In systems with large population imbalance, the minority atoms interact with the surrounding majority atoms and form dressed atomic states called polaron states. The binding energy of the polaron state has been measured by the radio frequency spectroscopy in a two-component Fermi system, and the transition from fermionic polarons to dimers has been observed as $a_s$ is varied~\cite{PhysRevLett.102.230402}. While single-polaron properties have been studied experimentally and found excellent agreement with theoretical predictions~\cite{PhysRevLett.98.180402,PhysRevLett.102.230402}, the fundamental understanding of effective interactions between the polarons is still lacking. The effective interactions between polarons are mediated by the surrounding fermions. This is a non-trivial many-body process, especially when the $s$-wave scattering length between the minority and majority atoms is resonantly large. 


The effective interactions is closely related to the stability of the system. For a system of two heavy and one light particles resonantly interacting with each other, the effective interaction between the heavy particles mediated by the light one is strongly attractive. Through this attraction, the Efimov states may appear, rendering the system unstable via the three-body loss~\cite{efimov1973energy,*braaten2006universality}. When the light fermions form the Fermi sea, on the other hand, the effective interaction is numerically found to acquire an additional repulsion, and the formation of the Efimov states is suppressed~\cite{PhysRevA.79.013629,PhysRevLett.106.145301}. This suggests that the Fermi sea tends to suppress the three-body loss. It is important to know under what condition such a suppression of the Efimov effect may occur, and whether a similar suppression can occur against four-body bound states, five-body bound states, etc.

In this paper, we study the effective interaction between polarons in a system of an arbitrary number of heavy particles immersed in the Fermi sea of light fermions. With the Born-Oppenheimer method, we analytically show that the effective interaction between the heavy particles vanishes in the limit of high light-fermion density. To be more specific, we prove the following theorem:

\noindent {\it Theorem.} Consider a mixture of $N$ heavy particles with mass $M$ and light degenerate fermions with mass $m$. The number of the light fermions is assumed to be so large that the Fermi sea is formed and the grand canonical ensemble can be applied to the light fermions. The statistics of the heavy particles is arbitrary: identical fermions, bosons, or distinguishable particles. The interaction between the heavy particles and the light fermions is assumed to be a zero-range interaction with the $s$-wave scattering length $a_s$~\cite{bethe1935quantum,*PhysRev.105.767}. The interaction between the light fermions and that between the heavy particles are assumed to be non-interacting. Then, within the Born-Oppenheimer method, the effective interaction $V_{\mathrm{eff}}(\bm{R_1}, \bm{R_2}, ..., \bm{R_{N}})$ between the $N$ heavy particles positioned at $(\bm{R_1}, \bm{R_2}, ..., \bm{R_{N}})$ mediated by the light fermions vanishes in the limit of $k_F \rightarrow +\infty$:
\begin{equation}
\lim_{k_F \rightarrow +\infty} V_{\mathrm{eff}}(\bm{R_1}, \bm{R_2}, ..., \bm{R_{N}}) =0,
\end{equation}
where $k_F$ is the Fermi momentum of the light fermions.

In other words, the effective interaction between the heavy polarons becomes small in the dense fermionic environment. Note that $a_s$ can take on any value as long as $\displaystyle k_F^{-1} \ll |a_s|, |\bm{R_{i}}-\bm{R_{j}}|$, so that the theorem is also applicable to the unitarity limit $1/a_s=0$.  This suggests that the formation of the $N$-body bound states associated with the Efimov effect is suppressed for any $N$ by a dense Fermi sea.



We ascribe this vanishing effective interaction to the screening in the neutral Fermi system. While the screening is a well-known phenomenon in the charged Fermi system, for a neutral Fermi system, little analytical results have been obtained in the resonantly interacting regime $ k_F |a_s| \gg 1$. The above theorem suggests that the screening phenomenon occurs for the neutral Fermi system, including the resonantly interacting regime.

This paper is organized as follows. In Sec. II, we define the effective interaction between the heavy particles with the Born-Oppenheimer method, and prove the main theorem. In \secref{disc}, we discuss the physical origin of the vanishing interaction and non-adiabatic effects. In Sec. IV, we conclude this paper.

\section{Mathematical description of the effective interaction}

\subsection{Definition of the effective interaction}

We define the effective interaction between the heavy particles $\displaystyle V_{\mathrm{eff}}(\bm{R_1}, \bm{R_2}, ..., \bm{R_{N}})$ in the same manner as in Refs.~\cite{PhysRevA.79.013629,PhysRevLett.106.145301}. With the Born-Oppenheimer approximation, we solve the Schr\"odinger equation for the light fermions by regarding the heavy particles as fixed impurities, positioned at $\bm{R_1}, \bm{R_2}, ..., \bm{R_{N}}$. Then, the obtained energy eigenvalue gives an effective interaction between the heavy particles. Since the light fermions are assumed not to interact with each other,  the solution of the Schr\"odinger equation for the light fermions is given by the Slater determinant $\displaystyle \Psi_L(\bm{r_1},\bm{r_2},...) = \mathcal{A}\prod_{i} \psi_{\bm{R}}^{(i)}(\bm{r_i})$, where $\mathcal{A}$ is the antisymmetrizer, $\bm{r_i}$'s are the positions of the $i$-th light fermions, and  $\displaystyle \psi_{\bm{R}}^{(i)}$'s are the solutions of the single-particle Schr\"odinger equation in the presence of the impurity potentials located at $\bm{R}=(\bm{R_1}, \bm{R_1}, ..., \bm{R_{N}})$. The energy eigenvalue for the light particle is the sum of the single-particle eigenvalues $\varepsilon_i(\bm{R})$ corresponding to $\displaystyle \psi_{\bm{R}}^{(i)}$: $\displaystyle  E(\bm{R})= \sum_{i} \varepsilon_i(\bm{R})$. The effective interaction between the heavy particles induced by the light fermions is obtained by subtracting the chemical potential of $N$ independent polarons:
\begin{equation}
V_{\mathrm{eff}}(\bm{R_1}, \bm{R_2}, ..., \bm{R_{N}})=E(\bm{R})- \lim_{|\bm{R_{ij}}|\rightarrow\infty}E(\bm{R}),
\end{equation}
where $\displaystyle \lim_{|\bm{R_{ij}}|\rightarrow\infty}$ means that all the heavy particles are far apart from each other so that they may be regarded as $N$ independent polarons.

For $N=2$ (i.e. two heavy particles immersed in the light Fermi sea), $\displaystyle V_{\mathrm{eff}}$ has been calculated and studied numerically~\cite{PhysRevA.79.013629}. As the number of the heavy particles increases, however, it becomes impractical to calculate $\displaystyle V_{\mathrm{eff}}$ numerically, since the effective interaction cannot be written as a simple sum of two-body interactions, but rather it includes all the three-body, four-body, ..., and $N$-body interactions. To circumvent this difficulty, we use a formal scattering theory to investigate $\displaystyle V_{\mathrm{eff}}$.



\subsection{\label{sec:Proof}Proof of the theorem}
To evaluate the effective interaction, we need the energy eigenvalues of the single-particle Schr\"odinger equation $\displaystyle \varepsilon_i(\bm{R})$ under the impurity potentials positioned at $\displaystyle \bm{R_1}, \bm{R_2}, ..., \bm{R_{N}}$. The total energy is the sum of the contributions from the continuum states ($\displaystyle \varepsilon_i \ge 0$) and the bound states ($\displaystyle \varepsilon_i < 0$):
\begin{equation}
\begin{split}
E(\bm{R})& = \sum_{\varepsilon_i \ge 0} \varepsilon_i(\bm{R})  + \sum_{\varepsilon_i <0} \varepsilon_i(\bm{R}).
\end{split}
\end{equation}
To evaluate the continuum part, we consider the scattering problem under the impurity potentials. Let us define the scattering phase shifts  $\delta_n(k)$ as eigenvalues of the S-matrix $\mathcal{S}(k)$~\cite{newton1977noncentral1,*newton1977noncentral2}:
\begin{equation}
\mathcal{S}(k)\bm{v}_n(k)= e^{2i \delta_n(k)}\bm{v}_n(k). \ \ 
\end{equation}
We first show that the continuum part of the effective interaction is related to the scattering phase shifts through the following lemma:

\noindent {\it Lemma }({\it generalized Fumi theorem})  
\begin{equation}
\begin{split}
\label{eq:Fumi_pot}V_{\mathrm{eff}}(\bm{R_1}, \bm{R_2}, ..., \bm{R_{N}}) &= V_{\mathrm{eff}}^{\mathrm{cont}}(\bm{R}) + V_{\mathrm{eff}}^{\mathrm{BS}}(\bm{R}),
\end{split}
\end{equation}
where $\displaystyle V_{\mathrm{eff}}^{\mathrm{cont}}(\bm{R})$ and $\displaystyle V_{\mathrm{eff}}^{\mathrm{BS}}(\bm{R})$ are the continuum and bound-state contributions, respectively, which are given by 
\begin{equation}
 V_{\mathrm{eff}}^{\mathrm{cont}}(\bm{R})=-\frac{1}{\pi m}\sum_n\int_0^{k_F}kdk  \delta_n(k) , 
\end{equation}
\begin{equation}
V_{\mathrm{eff}}^{\mathrm{BS}}(\bm{R}) =  \sum_{\varepsilon_i <0}\left[ \varepsilon_i(\bm{R}) -  \lim_{|\bm{R_{ij}}|\rightarrow\infty} \varepsilon_i(\bm{R}) \right].
\end{equation}
For the case of a single impurity ($N=1$) in the absence of bound states, this lemma reduces to the Fumi theorem~\cite{fumi1955}.

{\it Proof of lemma.}  We use the Friedel sum rule~\cite{friedel1952,PhysRev.121.1090,newton1977noncentral1,*newton1977noncentral2}
\begin{equation}
\label{eq:Fsum}N_I-N_0 = \frac{1}{\pi}\sum_n \delta_n(k_F),
\end{equation}
where $k_F$ is the Fermi momentum of the light fermions, and $N_I$ and $N_0$ are the numbers of the light fermions evaluated by the grand canonical ensemble with and without the impurity potential, respectively. For a central potential, the index $n$ represents the angular momentum quantum number $(l,m_l)$, and we recover the original Friedel sum rule. While the Friedel sum rule was originally proved for an ideal Fermi gas interacting with a central impurity potential~\cite{friedel1952}, it was subsequently generalized for an interacting system~\cite{PhysRev.121.1090} and also for a non-central potential~\cite{newton1977noncentral1,*newton1977noncentral2}. The impurity potential produced by $N$ heavy particles, in general, is a non-central potential. Even in such a case, the Friedel sum rule remains valid~\cite{newton1977noncentral1,*newton1977noncentral2}.

Using the Friedel sum rule, we can prove the generalized Fumi theorem. We recall that the number of the light fermions is related to the thermodynamic function $\Omega$ of the light fermions through the thermodynamic relation $\left(\dfrac{\partial \Omega}{\partial \mu}\right)_{T,V}=-N$. By integrating this relation with respect to the chemical potential for systems with and without the impurity potentials and using the Friedel's sum rule, we obtain
\begin{equation}
\begin{split}
\label{eq:Fumi}\Omega_I-\Omega_0 &=-\int_{\mu\ge 0} (N_I-N_0)d\mu - \int_{\mu < 0} (N_I-N_0) d\mu ,\\
&= -\frac{1}{\pi m}\sum_n\int_0^{k_F}kdk  \delta_n(k) -  \int_{\mu < 0} N_I  d\mu,\\
\end{split}
\end{equation}
where $\Omega_I$ and $\Omega_0$ are the thermodynamic function with and without the impurity potentials, respectively. In deriving the second equality, we use $N_0=0$ for $\mu<0$: there is no bound state in the absence of any potential. Because the light fermions are non-interacting, we can put $\mu_I\approx \mu_0 \approx \dfrac{k_F^2}{2m}$ in \equref{Fumi}. In fact, the shift in the energy level induced by the impurity potential is of the order of $V^{-\frac{1}{3}}$, where $V$ is the volume of the system, and the shift is negligible in the thermodynamic limit. Substituting $\displaystyle N_I=\sum_{i} \Theta (\mu - \varepsilon_i(\bm{R})) $, where $\Theta$ is the Heaviside step function, the second term in the second line of \equref{Fumi} can be evaluated as
\begin{equation}
 \int_{\mu < 0} N_I  d\mu = -\sum_{\varepsilon_i<0}\varepsilon_i(\bm{R})
 \end{equation}
The thermodynamic function of the light particles may be regarded as an effective interaction between the heavy particles, and we obtain
\begin{equation}
\begin{split}
\label{eq:Fumi2}E(\bm{R})= -\frac{1}{\pi m}\sum_n\int_0^{k_F}kdk  \delta_n(k) + V_{\mathrm{eff}}^{\mathrm{BS}}(\bm{R}) + \mathrm{Const}.\\
\end{split}
\end{equation}
We thus obtain \equref{Fumi_pot}.

{\it Proof of the theorem.} Using the lemma, we can prove the main theorem. We first use the following relation between the phase shifts and the Fredholm determinant $D(k)$~\cite{newton1977noncentral1,*newton1977noncentral2}:
\begin{equation}
\label{eq:delta_Det}\sum_n \delta_n(k) = -\frac{k}{4\pi}\int d\bm{x} U_{\bm{R}}(\bm{x}) + \frac{i}{2}\log\left[ \frac{D(k)}{D^*(k)}\right],
\end{equation}
where $U_{\bm{R}}$ is the sum of the impurity potentials produced by the heavy particles. The Fredholm determinant is defined from the kernel matrix $\mathcal{K}(k)$ of the Lippmann-Schwinger equation as $D(k)=\det[1-\lambda \mathcal{K}(k)]_{\lambda=1}$, and has the following properties for a short-ranged, non-central potential~\cite{newton1977noncentral1,*newton1977noncentral2}: 
\begin{enumerate}
\item $D(k)$ is well-defined and analytic for $\mathrm{Im}k \ge 0$; 
\item $\lim_{|k|\rightarrow \infty} D(k)=1$ for $\mathrm{Im}k\ge 0$; 
\item For a real $k$, $D^*(k)=D(-k)$; 
\item The zero of $D(k=i\kappa)=0$ in the upper-half complex $k$-plane has a one-to-one correspondence with a bound state with its energy $\varepsilon= -\frac{\kappa^2}{2m}$;
\item The zero of $D(k)$ can appear either on a positive imaginary plane, or at the origin $k=0$ for a short-range potential.
\end{enumerate}

Substituting \equref{delta_Det} into $V_{\mathrm{eff}}^{\mathrm{cont}}$ and differentiating both sides with respect to $\bm{R}_i$, we find
\begin{equation}
\begin{split}
\label{eq:del_Veff} \nabla_{\bm{R_i}}V_{\mathrm{eff}}^{\mathrm{cont}}&=-\frac{i}{2\pi m}\int_{0}^{k_F}kdk  \nabla_{\bm{R_i}}\log\left[ \frac{D(k)}{D^*(k)}\right]\\
 &=-\frac{i}{2\pi m}\int_{-k_F}^{k_F}kdk  \frac{\nabla_{\bm{R_i}}D(k)}{D(k)}. 
 \end{split}
\end{equation}
In deriving the second equality, we have used the property 3 to transform the integration of $D^*(k)$ into that of $D(k)$ along the negative real axis. Now, let us take the limit $k_F\rightarrow +\infty$. The properties 2 and 3 ensure that there is a well-defined limit for \equref{del_Veff}. Furthermore, the properties 1-3 also justify the change of the integration contour into paths $C_j$'s encircling the zeros of $D(k)$:
\begin{equation}
\label{eq:nabla_last}\nabla_{\bm{R_i}}V_{\mathrm{eff}}^{\mathrm{cont}}=-\frac{i}{2\pi m}\sum_j \int_{C_j}kdk  \frac{\nabla_{\bm{R_i}}D(k)}{D(k)}. 
\end{equation}
Close to the zero point of $k\approx i\kappa_j$, we can put $D(k) =  \alpha_j (k-i\kappa_j) + O((k-i\kappa_j)^2)$ if the bound state is not degenerate~\cite{newton1977noncentral1,*newton1977noncentral2}. Substituting this into \equref{nabla_last} and performing the integration, we obtain 
\begin{equation}
\begin{split}
\label{eq:main_res}\nabla_{\bm{R_i}}V_{\mathrm{eff}}^{\mathrm{cont}}& =\nabla_{\bm{R_i}}\sum_j \frac{\kappa_j^2}{2m}\\
& = - \nabla_{\bm{R_i}} V_{\mathrm{eff}}^{\mathrm{BS}} (\bm{R}).
\end{split}
\end{equation}
When the bound states are $n$-fold degenerate, $D(k)$ behaves as $D(k)\approx \alpha(k-i\kappa_j)^n + O((k-i\kappa_j)^{n+1})$~\cite{newton1977noncentral1,*newton1977noncentral2}. Even with such a degeneracy, we can derive \equref{main_res} from \equref{nabla_last}, and the above result remains valid. Thus, the continuum contribution exactly cancels the bound-state one, and the effective interaction vanishes.

\section{\label{sec:disc}Physical origin of the vanishing effective interaction and non-adiabatic effects}

\begin{figure}
 \includegraphics[width=9cm,clip]{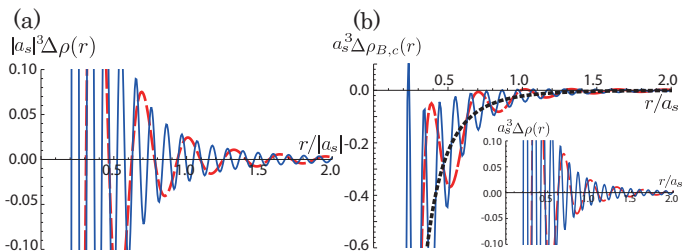}%
 \caption{\label{fig:dens_plot}(Color online) Density variations induced by a single impurity for (a) $a_s<0$, and (b) $a_s>0$. In (a), $\Delta \rho(r)$ is shown for $k_F a_s =-10$ (red dashed curve) and $k_F a_s =-30$ (blue solid curve). In (b), $\Delta \rho_c(r)$ is shown for $k_F a_s =10$ (red dashed curve) and $k_F a_s =30$ (blue solid curve). The black dotted curve represents $-\Delta \rho_B(r)$. The inset shows the total density variation $\Delta \rho(r)=\Delta \rho_B(r)+\Delta \rho_c(r)$.}
 \end{figure}

The vanishing effective interaction is closely related to the behavior of the density variation caused by the impurities. To see this point, let us consider a single impurity immersed in the light Fermi sea. The density variation $\Delta \rho (r)$ induced by the impurity can be expressed as the sum of the bound-state and continuum-state contributions $\Delta \rho (r)=\Delta \rho_B (r)+\Delta \rho_c (r)$, where
\begin{equation}
\begin{split}
\Delta \rho_B(r) &= \frac{1}{2\pi r^2}\theta(a_s)\frac{e^{-\frac{2r}{a_s}}}{a_s},\\
\Delta \rho_c(r) &= \frac{1}{2\pi^2 r^2}\int_0^{k_F}dk [\sin^2(kr+\delta_0(k))- \sin^2 kr],
\end{split}
\end{equation}
and $\delta_0(k)$ is the $s$-wave phase shift induced by the impurity: $\tan \delta_0(k)=-ka_s$. In \figref{dens_plot}, we show $\Delta \rho (r)$'s for several values of $k_F a_s$. For $a_s<0$ (\figref{dens_plot} (a)), there is no bound state and $\Delta \rho_B(r)=0$. The density variation reflects the Friedel oscillations characterized by $k_F$. As $k_F$ increases, the oscillations become faster, but the amplitude remains the same. For $a_s>0$ (\figref{dens_plot} (b)), $\Delta \rho (r)$ is the sum of $\Delta \rho_B(r)$ and $\Delta \rho_c(r) $. We note that $\Delta \rho_c(r)$ follows the $-\Delta \rho_B(r)$ curve on average, so that the continuum states screen the bound-state contribution. Due to this screening effect, $\Delta \rho (r)$ undergoes fast oscillations around the zero on both positive and negative sides of $a_s$. Thus, the interaction between the impurities mediated by the density fluctuation should become weaker as $k_F$ increases regardless of the value of the $s$-wave scattering length or the distance between the heavy particles.

The screening in the neutral Fermi system is qualitatively different from that in the charged Fermi system. In the former case, the induced interaction vanishes due to the cancellation of the bound-state contribution by the continuum one. If we add the Hamiltonian with a direct heavy-heavy interaction, the direct interaction is not screened and the effective interaction between the heavy particles remains finite. On the other hand, in the charged Fermi system, the direct interaction between the heavy particles is canceled by the induced one.

For the effective interaction between two heavy particles ($N=2$), the exact cancellation of the effective interaction can also be shown by explicit calculation of the subleading contribution to the effective interaction at large density. When $k_F$ is large, one can show that $V_{\mathrm{eff}}$ behaves as
\begin{equation}
\begin{split}
\label{eq:correct}V_{\mathrm{eff}}(r) =&-\frac{\cos 2k_F r}{2\pi m  k_F r^3}-\frac{\sin 2k_F r}{4\pi m k_F^2 r^4}  + \frac{\sin 2k_F r}{\pi m r^3 k_F ^2a_s}  \\
&+ \frac{1}{m}O\left( \frac{1}{k_F^3 a_s r^4}, \frac{1}{k_F^3 a_s^3r^2},\frac{1}{k_F^3 a_s^5},... \right).
\end{split}
\end{equation}
The effective interaction is suppressed by a factor of $k_F^{-1}$, and becomes small as $k_F$ increases.


Our theorem suggests that the formation of the $N$-body bound states associated with the Efimov effect is suppressed for any number of heavy particles by the dense Fermi sea. For two heavy particles $N=2$, the suppression of the Efimov effect was numerically found in Ref.~\cite{PhysRevLett.106.145301}. Our work presents a more general argument, which is applicable for arbitrary number of heavy particles. Since the appearance of the Efimov associated bound states is closely related to the loss processes in ultracold atom experiments, this implies that the $N$-body losses in a resonantly interacting heavy-light mixture may be suppressed by the Fermi sea effects for any $N$.

The non-adiabatic effects beyond the Born-Oppenheimer method need to be considered in discussing the dynamics of the polarons. For systems with moderate mass imbalance, such as a  $^{40}$K-$^{6}$Li mixture~\cite{kohstall2012metastability}, the Born-Oppenheimer approach fails. Even for mixtures with extreme mass imbalance recently realized, such as $^{133}$Cs-$^6$Li~\cite{PhysRevA.87.010701,*PhysRevA.87.010702,PhysRevA.87.010701,*PhysRevA.87.010702} or $^{173}$Yb-$^6$Li~\cite{PhysRevLett.106.205304,*PhysRevA.84.011606}, the non-adiabatic corrections may affect the dynamics significantly. The dynamics of a single heavy particle in a fermionic environment has been studied in the absorption spectra of X-ray~\cite{PhysRev.163.612,*PhysRev.178.1097} or the muon diffusion~\cite{Kondo_Yoshimori201112} in metals. It has been found that the motion of heavy particles create particle-hole excitations in the Fermi sea, which leads to dissipation~\cite{rosch1999quantum}. It has also been suggested that the non-adiabatic effects can create non-trivial correlation between heavy particles~\cite{PhysRevB.11.2122,*PhysRevLett.97.250601}. Our theorem suggests that in the presence of a dense Fermi sea, the adiabatic contribution becomes so small that the dynamics of the polarons is governed by the non-adiabatic contributions. Whether such non-adiabatic corrections remain significant or become negligibly small in the high-density limit remains to be clarified..

\section{Conclusion} 

With the Born-Oppenheimer method, we have proved that the effective interaction between an arbitrary number of heavy polarons mediated by light fermions vanishes in the limit of high fermion density. Our theorem holds for any value of the $s$-wave scattering length including the unitarity regime. We ascribe the vanishing effective interaction to the screening effect in the neutral Fermi system. Our work suggests that the $N$-body Efimov effect is suppressed in the presence of a sufficiently dense Fermi sea of light particles. This implies that the $N$-body loss processes may be suppressed by a dense Fermi sea.

\begin{acknowledgments}
We thank Y. Nishida, P. Naidon, and D. Blume for illuminating discussions. This work was supported by KAKENHI 22340114, 
a Grant-in-Aid for Scientific Research on Innovation Areas ``Topological Quantum Phenomena" (KAKENHI 22103005),
the Global COE Program "the Physical Sciences Frontier,"
and the Photon Frontier Network Program,
from MEXT of Japan. S. E. acknowledges support from JSPS (Grant No. 237049).
\end{acknowledgments}

%

\end{document}